\documentclass[10pt,amssymb,superscriptaddress,showkeys,twocolumn,prl]{revtex4-2}
\usepackage{graphicx}
\usepackage{dcolumn}
\usepackage{bm}
\usepackage{changes}
\usepackage{amsmath}
\usepackage{amssymb}
\usepackage{titlesec}
\usepackage{lineno}
\usepackage{hyperref}

\makeatletter
\renewcommand{\@caption@fignum@sep}{\space} 
\makeatother
\titleformat{\section}[block]{\normalfont\bfseries\large}{\thesection}{1em}{}
\titlespacing*{\section}{0pt}{\baselineskip}{0.5\baselineskip}

\titleformat{\subsection}[block]{\normalfont\bfseries\small}{\thesubsection}{1em}{}
\titlespacing*{\subsection}{0pt}{\baselineskip}{0.5\baselineskip}

\hypersetup{
 colorlinks=true,
 linkcolor=blue,
 filecolor=blue,  
 urlcolor=blue,
 citecolor=blue,
}

\begin{document}

\title{Stable high-charge vortex dissipative solitons in azimuthally modulated waveguide arrays with localized gain}

\author{Changming Huang}
\email{hcm123\_2004@126.com}
\affiliation{Department of Physics, Changzhi University, Changzhi, Shanxi 046011, China}

\author{Qidong Fu}
\affiliation{School of Physics and Astronomy, Shanghai Jiao Tong University, Shanghai 200240, China}

\author{Li Ma}
\affiliation{Department of Physics, Changzhi University, Changzhi, Shanxi 046011, China}

\date{\today}

\begin{abstract}
We study the existence and dynamical properties of vortex solitons in Kerr media supported by azimuthally modulated waveguide lattices with localized gain and nonlinear loss. 
In this dissipative system, we find that the accessible topological charge of vortex solitons is strongly determined by the number of waveguide channels, with higher-order charges requiring progressively larger arrays. Power curves of vortex solitons with different charges exhibit clear separation in large arrays but become less distinguishable in smaller ones. Furthermore, these robust vortex solitons can be excited with nearly vanishing power thresholds, and higher-charge vortices display enhanced propagation stability compared with lower-charge states. These findings expand the family of dissipative vortex solitons supported by waveguide lattices and provide a route to the realization of stable high-symmetry vortex states.
\end{abstract}

\maketitle

\section{Introduction}
Optical vortices, characterized by phase singularities and orbital angular momentum, have emerged as a central topic in modern optics \cite{PhysRevE.64.026601,desyatnikov2005optical,WangNanophotonics2018,shen2019optical,MALOMED2022112526}.
Their unique properties underpin a broad spectrum of applications, spanning optical tweezers \cite{Gahagan:96,Simpson:97}, optical spectroscopy \cite{Furhapter:05}, and quantum information processing \cite{mair2001entanglement,Gibson:04}.
Vortex beams can, in principle, be constructed through the superposition of degenerate higher-order modes \cite{Huang:16,huang2024vortex,KARTASHOV2023113919}, and have been generated experimentally using a variety of techniques, including mode converters employing astigmatic lenses \cite{BEIJERSBERGEN1993123,PhysRevA.54.R3742}, spiral phase plates \cite{BEIJERSBERGEN1994321}, twisted elliptical fibers \cite{science.1097631}, and photonic crystal fibers \cite{science.1223824,Xi:14}, among other platforms \cite{Heckenberg:92,Huang2015}.

In conservative systems, where gain and loss are absent, the propagation dynamics of optical vortices have been extensively investigated, both with and without optical lattices \cite{malomed2002stability,PhysRevLett.69.2503,Yang:03,PhysRevLett.93.063901,PhysRevLett.95.203904,Kartashov:06,PhysRevLett.129.123903}.
In such settings, stability emerges only through a delicate balance among diffraction, lattice effects, and nonlinearity.

Dissipative systems represent the most general case, where the sustained and stable propagation of light beams results from additional gain-loss balances between linear (or nonlinear) gain and nonlinear (or linear) losses.
Such interactions underlie the existence of dissipative vortex solitons, which have been extensively investigated in diverse contexts. Examples include systems governed by the Ginzburg-Landau equation \cite{RevModPhys.74.99,PhysRevLett.96.013903,PhysRevE.63.016605,soto2009dissipative,PhysRevA.80.033835,PhysRevA.82.023813,PhysRevLett.105.213901,PhysRevA.75.033811,PhysRevA.76.045803,PhysRevLett.97.073904}, mode-locked lasers \cite{grelu2012dissipative}, and media featuring localized gain combined with linear background losses \cite{Huang:OL13,PhysRevLett.132.213802}, as well as those with nonlinear absorption \cite{Kartashov:OL10,Lobanov:OL2011,Borovkova:OL11,Borovkova:OL11X,lobanov2022fundamental}.


The balances between linear localized gain and nonlinear loss offers greater flexibility than parity-time symmetric systems, which require globally antisymmetric gain-loss distributions \cite{Kartashov:16,PhysRevA.95.053613,Leykam:13,XU2024115043,xu2024vortex,zhao2025vortex}. In such dissipative systems, studies have largely focused on settings without optical lattices. Apart from reference \cite{Kartashov:OL10}, localized ring- or necklace-shaped vortex solitons are confined to the regions of localized gain and generally exist only above a certain gain threshold \cite{Lobanov:OL2011,Borovkova:OL11,Borovkova:OL11X,lobanov2022fundamental}.
To the best of our knowledge, these works have been restricted to vortex solitons with topological charges $m\leq3$, while the properties of dissipative higher-order vortices with $m\geq4$ remain unexplored.
Moreover, in the absence of optical lattices, necklacelike dissipative vortex solitons tend to develop asymmetric profiles at elevated gain levels \cite{Lobanov:OL2011}. This naturally raises the question of whether optical lattices can preserve vortex symmetry, and how the optical properties of higher-charge dissipative vortices are modified in the presence of lattice potentials? Thus, addressing these questions is the central aim of this work.

Here, we investigate dissipative vortex solitons in Kerr media featuring azimuthally modulated waveguide arrays with localized gain and nonlinear loss. 
Both the existence and propagation for high-charge vortices are considered.
We find that such robust vortex solitons can be excited with nearly vanishing power thresholds, that the number of azimuthal waveguides strongly constrains the maximum sustainable topological charge, and that higher-charge vortex solitons are more readily stabilized than their lower-charge counterparts.
These findings open new perspectives for the stabilization of high-charge dissipative vortex solitons.

\section{Theoretical model}
We consider the propagation of laser beam
in a two-dimensional optical lattice with 
a localized linear gain landscape and nonlinear absorption, governed by the dimensionless nonlinear
Schr\"{o}dinger equation:
\begin{equation}\label{eq1}
	i\frac{\partial \Psi}{\partial z} =-\frac{1}{2}\left(\frac{\partial^2\Psi}{\partial x^2}+\frac{\partial^2\Psi}{\partial y^2}\right)-V\Psi-(\sigma+i\kappa) \left |  \Psi\right |^2\Psi,  
\end{equation}
Here, $\Psi(x,y,z)$ denotes the amplitude of the light field, 
and $x$, $y$ and $z$ are the normalized transverse and longitudinal coordinates, respectively;
parameter $\sigma=+1$ corresponds to focusing nonlinearity,
and $\kappa>0$ quantifies the nonlinear loss coefficient.
The function $V(x,y)=(p_r-ip_i)V(x,y)$ is complex, with its corresponding physical quantities  $V_r(x,y)= p_rV(x,y)$ representing a two-dimensional optical lattice distribution and $V_i(x,y)=p_iV(x,y)$ describing a linear localized gain profile, both exhibiting similar spatial patterns, here, $p_r$ and $p_i$ denote the depth of the optical lattice and the strength of the linear gain, respectively. 
We propose that $V(x,y)$ corresponds to a ring-shaped Gaussian waveguide array, which can be mathematically expressed as 
$V(x,y)=\sum_{k=1}^{n}\exp\left[-(x-r_c\cos \varphi_k)^2/d^2-(y-r_c\sin \varphi_k)^2/d^2\right]$, 
here $\varphi_k=2\pi(k-1)/n$, $n\in \mathbb{Z}$ is the number of waveguides,
$r_c$ and $d$ correspond to the radius of the ring-shaped waveguide array and the width of the Gaussian waveguides, respectively.

We search for the profiles of vortex solitons numerically in the form $\Psi(x,y,z)=\phi(x,y)\exp(i\beta z)=[\phi_r(x,y)+i\phi_i(x,y)]\exp(i\beta z)$, where  $\phi$, $\phi_r$ and $\phi_i$ are the complex, real and imaginary parts of the phase-containing profiles, respectively, and $\beta$ represents the propagation constant.
Substituting this expression into Eq. (\ref{eq1}), we derive a set of coupled equations involving the real and imaginary components, $\phi_r$ and $\phi_i$:
\begin{eqnarray}
	\frac{1}{2}\left(\frac{\partial^2 \phi_{r}}{\partial x^2}+\frac{\partial^2 \phi_{r}}{\partial y^2}\right)+(V_{r}-\beta) \phi_{r}+V_{i} \phi_{i}+f_1=0,& \label{eq2} \\ 
	\frac{1}{2}\left(\frac{\partial^2 \phi_{i}}{\partial x^2}+\frac{\partial^2 \phi_{i}}{\partial y^2}\right)+(V_{r}-\beta)\phi_{i}-V_{i} \phi_{r}+f_2=0.& \label{eq3}
\end{eqnarray}
Here, $f_1=\sigma(\phi_{r}^{3}+\phi_{i}^{2} \phi_{r})-\kappa(\phi_{r}^{2} \phi_{i}+\phi_{i}^{3})$, 
$f_2=\sigma(\phi_{r}^{2} \phi_{i}+\phi_{i}^{3})+\kappa(\phi_{r}^{3}+\phi_{i}^{2} \phi_{r})$.
Eqs.~(\ref{eq2}) and (\ref{eq3}) 
can be solved numerically using Newton's iteration method. The power of a vortex soliton is quantified by the integrated formula: $U=\iint_{-\infty}^{+\infty}(|\phi_r|^2+|\phi_i|^2)dxdy$,
and the characteristic phase of vortex solitons is defined by their topological charge: 
$m=\frac{1}{2\pi}\text{Im}\int_0^{2\pi}\partial\phi(r_0,\varphi)/\partial\varphi/\phi(r_0,\varphi)d\varphi$,
where $r_0$ denotes the radius of a fixed circular path
and $\varphi$ represents the azimuthal coordinate.

\begin{figure*}
	\centering
	\includegraphics[width=1.36\columnwidth]{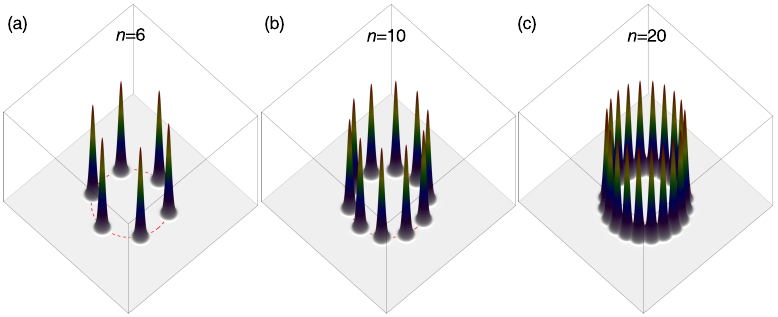}
	\caption{Representative distributions of ring-shaped waveguide arrays.
		Waveguide indices for (a) $n = 6$ , (b) $n = 10$, and (c) $n = 20$. The red dashed line marks the ring at $r_c = 5.25$. The optical lattice and linear localized gain exhibit similar distributions. $x,y\in [-12,+12]$ in all panels.}
	\label{fig1}
\end{figure*}

In the following discussion, we set $p_r=7$, $\kappa=1$, $r_c=5.25$ and $d=0.5$, while varying $n$ and $p_i$.
Typical optical lattices with varying indices $n$ are illustrated in Fig.~\ref{fig1}. Here, the optical waveguide array is arranged along a ring of fixed radius, with the spacing between adjacent waveguides decreasing as $n$ increases. This design contrasts with previously reported azimuthally modulated lattices, in which the ring radius itself changes with the number of waveguides \cite{kartashov2005soliton,ZhengJOSAB:11}.



\begin{figure*}
	\centering
	\includegraphics[width=1.78\columnwidth]{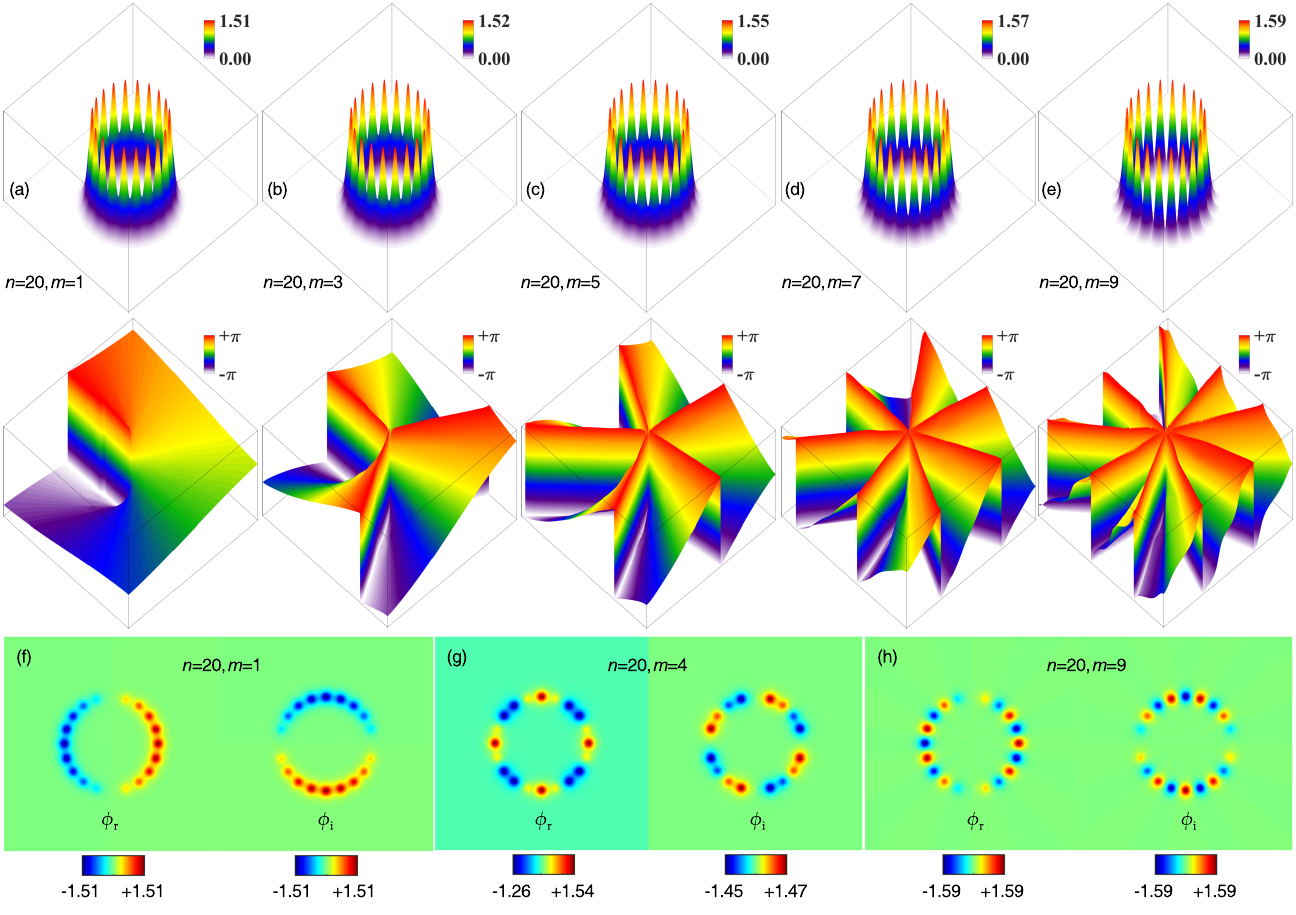}
	\caption{\textcolor{black}{Typical field modulus $|\phi|$ and phase distributions (a-e), and real and imaginary parts $\phi_\text{r,i}$ (f-h) of vortex solitons. (a,f) $m=1$; (b) $m=3$; (c) $m=5$; (d) $m=7$; (e,h) $m=9$; (g) $m=4$. $n=20$, $p_i=2$ and $x,y\in [-12,+12]$ in all panels.}
	}
	\label{fig2}
\end{figure*}
\begin{figure*}
	\centering
	\includegraphics[width=1.38\columnwidth]{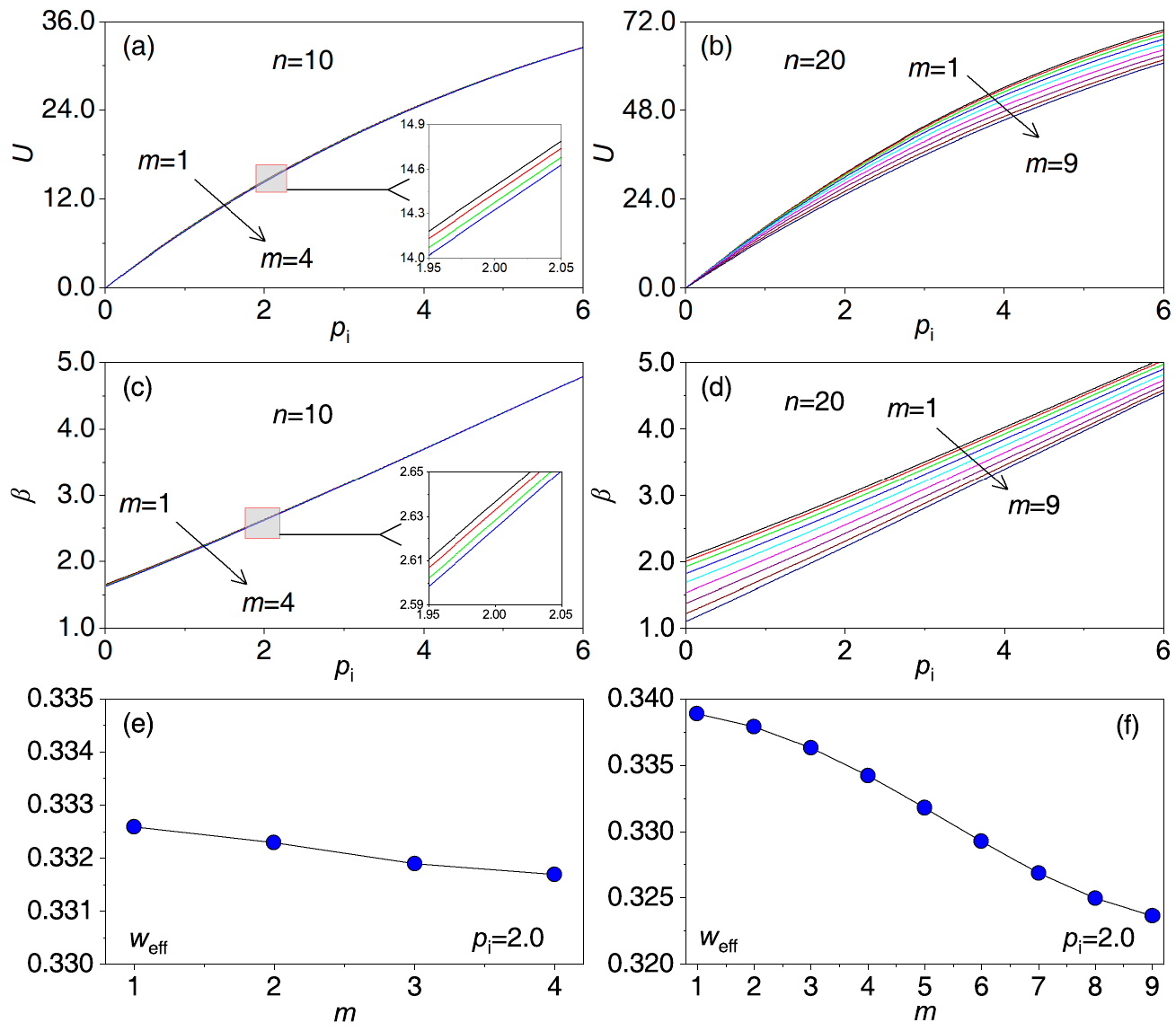}
	\caption{Power $U$, propagation constant $\beta$ and effective width $w_\text{eff}$ of vortex solitons with different topological charges $m$ under varying linear localized gain levels $p_\text{im}$. Panels (a), (c), and (e) correspond to $n=10$, while (b), (d), and (f) display results for $n=20$. Arrows in panels (a-d) indicate the direction of increasing $m$. The insets in (a) and (c) show magnified views of the gray-shaded regions.}
	\label{fig3}
\end{figure*}
\begin{figure*}
	\centering
	\includegraphics[width=1.38\columnwidth]{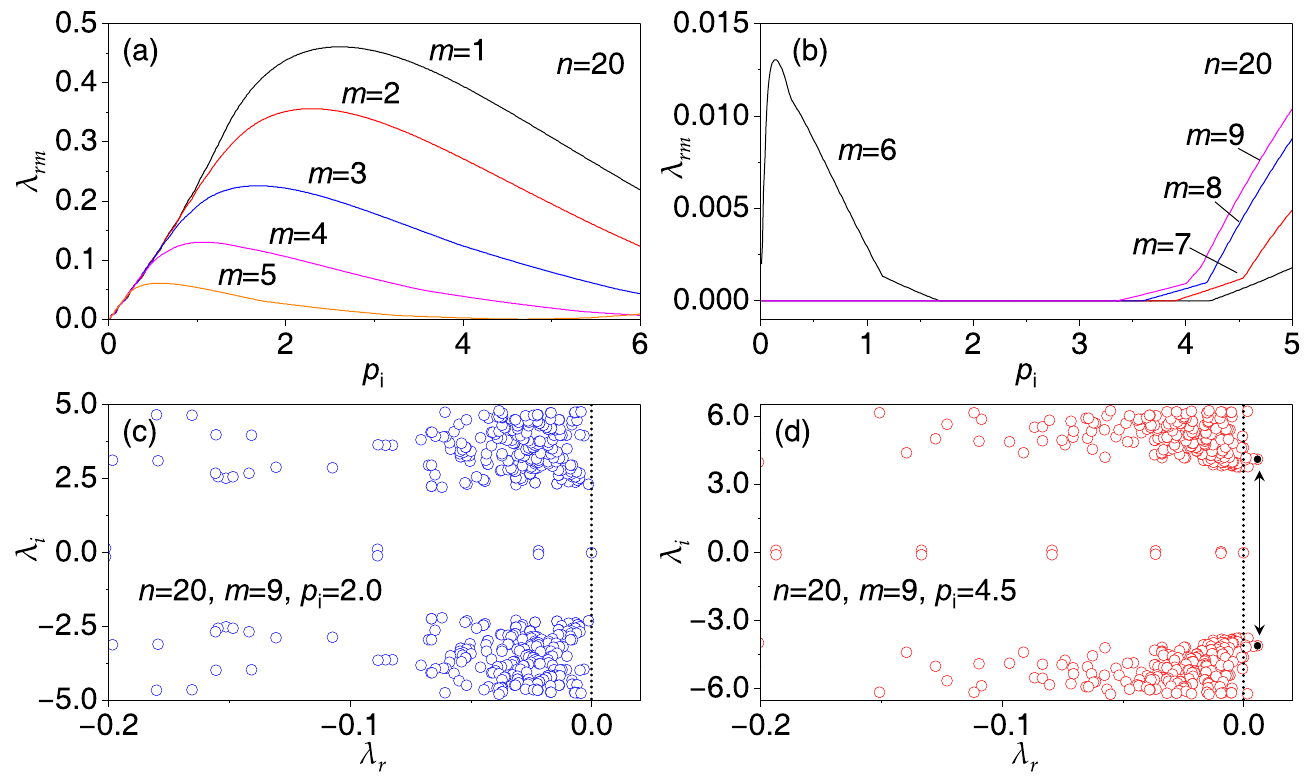}
	\caption{Dependence of maximum instability growth rate $\lambda_{rm}$ on gain parameter $p_i$ for distinct topological charges. (a) $m=1\sim 5$, and (b) $m=6\sim 9$. Eigenvalue spectra of (c) stable ($p_i=2.0$) and (d) unstable ($p_i=4.5$) vortex solitons with topological charge $m=9$. 
		\textcolor{black}{In panel (d), the maximum $\lambda_r$ is marked by black solid markers and annotated with bidirectional arrows.}
		The dashed black lines in (c) and (d) indicate $\lambda_{r}=0$.
		$n=20$ in all panels.
	}
	\label{fig4}
\end{figure*}
\begin{figure*}
	\centering
	\includegraphics[width=1.38\columnwidth]{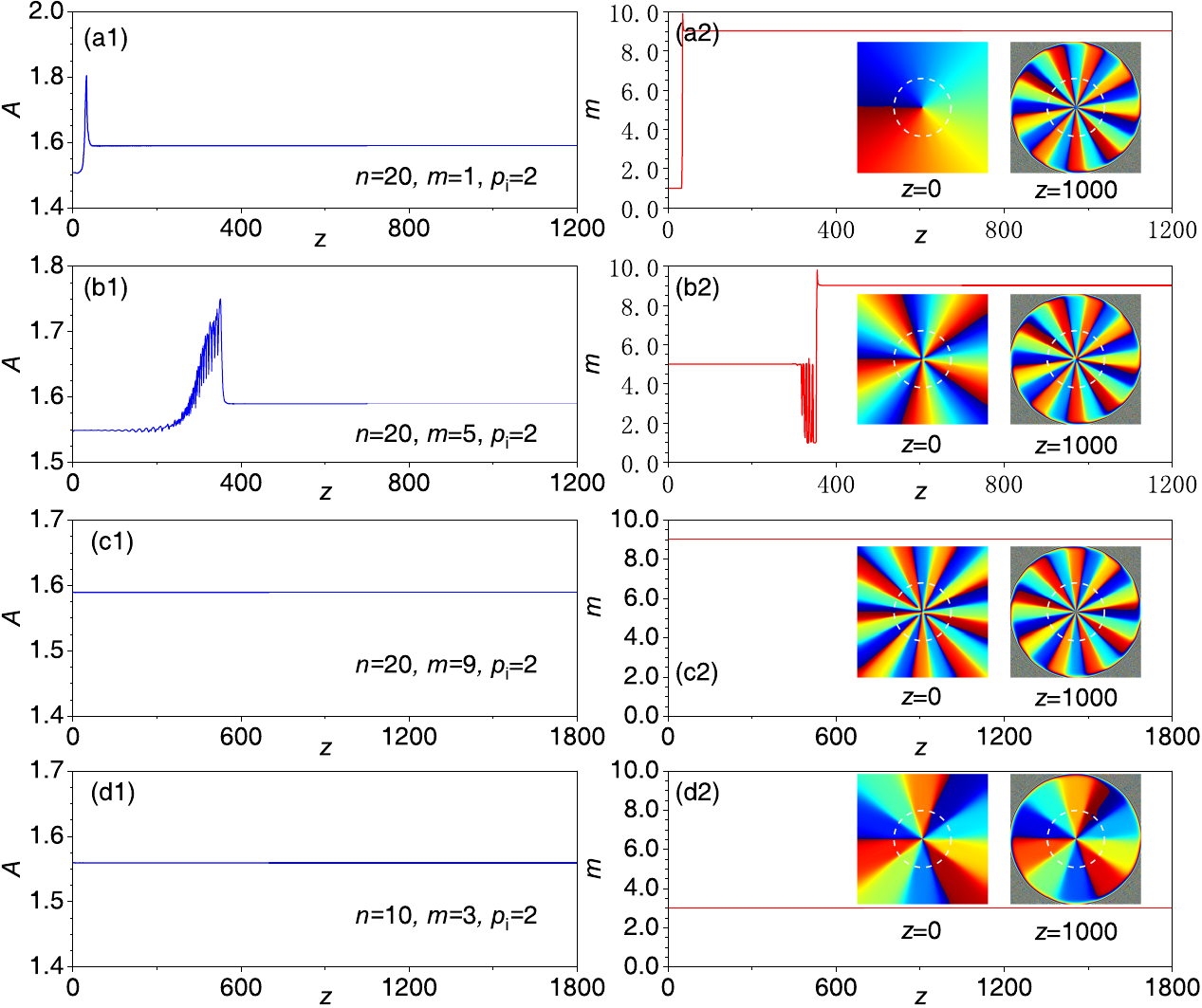}
	\caption{Evolutions of amplitude $A$ (left panels) and topological charge $m$ (right panels) of dissipative vortices versus propagation distance $z$.
		Rows $1\sim2$ display unstable propagation dynamics, while rows $3\sim4$ exhibit stable evolutions. 
		Parameter configurations: $n=20,m=1,p_i=2$ in (a1) and (a2); $n=20,m=5,p_i=2$ in (b1) and (b2); $n=20,m=9,p_i=2$ in (c1) and (c2); and $n=10,m=3,p_i=2$ in (d1) and (d2). Insets in right panels show phase distributions at $z=0$ and $z=1000$.
	}
	\label{fig5}
\end{figure*}
\begin{figure*}
	\centering
	\includegraphics[width=1.38\columnwidth]{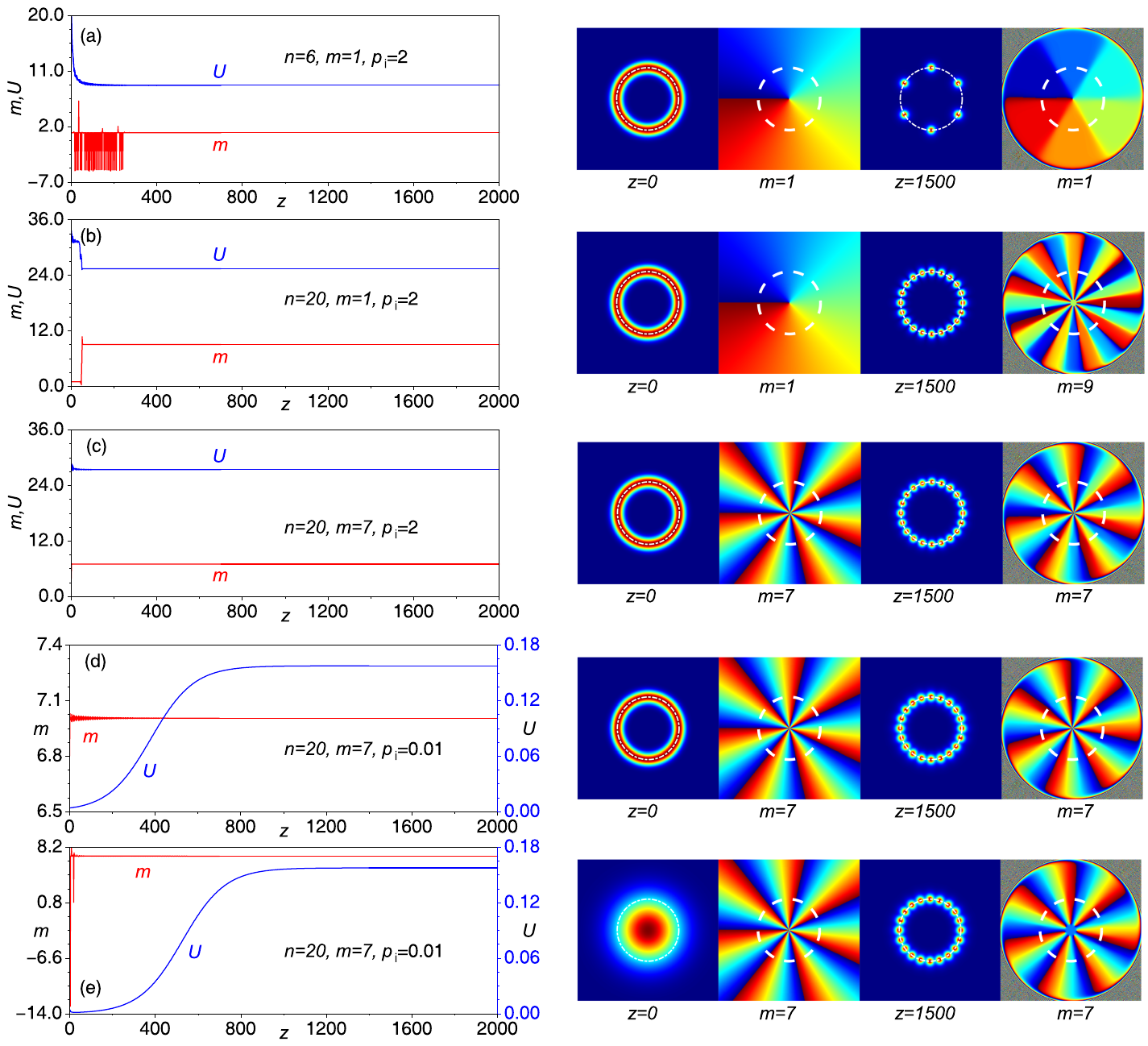}
	\caption{Excitation examples of dissipative vortex solitons with different topological charges.
		(a) $n=6,m=1$; (b) $n=20,m=1$; (c-e) $n=20,m=7$. 
		\textcolor{black}{$r_0=5.25$ and $w=1$ in (a-d), $r_0=0$ and $w=5.25$ in (e).}
		$p_i=2.0$ and $A_{in}=1.0$ in (a), (b) and (c), $p_i=0.01$ and $A_{in}=0.01$ in (d) and (e).
	}
	\label{fig6}
\end{figure*}

\section{Results and discussion}
Representative field-modulus and phase distributions of vortex solitons with varying topological charges are shown in Fig.~\ref{fig2}. 
As can be seen, the light spots of vortex solitons are uniformly confined within the azimuthally modulated waveguides, precisely overlapping the regions of localized gain \textcolor{black}{[upper panels of Figs.~\ref{fig2}(a-e)]}.
At fixed gain levels, the amplitude of the vortex soliton increases monotonically with topological charge $m$, and the phase varies by $m\times2\pi$ along the annular waveguide array for different topological charges \textcolor{black}{[lower panels of Figs.~\ref{fig2}(a-e)]}.
\textcolor{black}{In addition, Figs.~\ref{fig2}(f-h) show the real and imaginary parts of representative dissipative vortex solitons. For low-charge states (e.g., $m=1$), the field peaks in both components appear several nearly in phase at adjacent sites [Fig.~\ref{fig2}(f)]. With increasing $m$, the number of such in-phase neighbors decreases [Fig.~\ref{fig2}(g)]. At higher charges (e.g., $m=9$), the corresponding states develop staggered, out-of-phase patterns across adjacent sites [Fig.~\ref{fig2}(h)].}
Owing to the presence of the waveguide lattice, the bright spots within the vortex soliton preserve their symmetric arrangement even at elevated gain levels, in sharp contrast to highly asymmetric spatial profiles observed for vortex solitons in the absence of an optical lattice \cite{Lobanov:OL2011}.

The existence domains of vortex solitons with different topological charges are systematically investigated 
in different waveguide configurations.
The power $U$ and propagation constant $\beta$ variations of vortex solitons with different topological charges $m$ under varying linear localized gain levels $p_{i}$ are shown in Fig.~\ref{fig3}.
Representative cases for indices $n=10$ and $n=20$ are presented to illustrate the general behavior.
Four key features emerge: 
(i) For a fixed topological charge $m$, the soliton power $U$ (or propagation constant $\beta$) 
increases monotonically with the linear gain amplitude $p_{i}$ [Figs.~\ref{fig3}(a-d)]. 
\textcolor{black}{As the gain amplitude $p_{i}$ tends to zero, the power of dissipative vortex solitons vanishes, and their propagation constants approach those of the linear modes in the conservative ring-shaped array.
	Specifically, the propagation constant of an $m$-charge vortex soliton approaches the eigenvalues of the degenerate modes with indices $2m$ and $2m+1$ [Figs.~\ref{fig3}(c) and \ref{fig3}(d)].}
(ii) At lower waveguide indices ($n\leq10$), the ($p_{i},U$) or ($p_{i},\beta$) curves of vortex solitons with different charges $m$ exhibit minimal separation.
\textcolor{black}{This feature arises from weak coupling between well-separated waveguides at the fixed ring radius, resulting in a very small splitting of powers and propagation constants.}
(iii) Vortex solitons with higher-order topological charges exhibit an overall reduction in power compared with their lower-charge counterparts, particularly evident in the $n=20$ configuration. Vortex solitons exhibit a similar functional dependence in their propagation constants.
(iv) For a given waveguide number $n$, the structure possesses $C_n$ rotational symmetry. The maximum attainable topological charge 
$m$ of vortex solitons is fundamentally constrained by this symmetry order $n$.
For structures with odd symmetry orders ($n\geq 3$), the maximum charge follows $m=(n-1)/2$, while for even orders ($n\geq 4$), the limit becomes $m=(n-2)/2$, a result that has also been demonstrated in conservative systems \cite{kartashov2005soliton,ZhengJOSAB:11}.

For a fixed gain parameter (e.g., $p_i=2.0$), the field modulus amplitude exhibits a monotonic increase with the topological charge $m$ [see the top panels of Fig.~\ref{fig2}], while its corresponding power decreases gradually [Fig.~\ref{fig3}(b)]. 
These observations indicate that, at a fixed gain, dissipative vortex solitons in azimuthally modulated waveguide arrays undergo a progressive narrowing of their width as the topological charge increases. The effective width of solitons can be defined as $w_\text{eff}=[\int_{0}^{\infty}(x-x_c)^2| \phi(x,y=0)|^2dx/\int_{0}^{\infty}| \phi(x,y=0)|^2dx]^{1/2}$,
with $x_c=\int_{0}^{\infty}x|\phi(x,y=0)|^2dx/\int_{0}^{\infty}|\phi(x,y=0)|^2dx$. 
\textcolor{black}{To quantitatively validate this optical phenomenon, we systematically analyze the dependence of $w_\text{eff}(m)$ [Figs.~\ref{fig3}(e) and \ref{fig3}(f)]. The results show that the width of dissipative vortex solitons confined in the ring-shaped waveguide array indeed decreases with increasing $m$, regardless of whether the number of waveguides is large or small.}

The stability of vortex solitons with different topological charges supported by this azimuthally modulated structure constitutes a key focus of our investigation. To elucidate this optical property,
we conduct linear stability analysis of the stationary solutions by introducing perturbations of the form:
$\Psi(x,y,z)=[\phi_r+i\phi_i+(g_1+ig_2)e^{\lambda z}]e^{i\beta z}$, 
where $\lambda=\lambda_r+i\lambda_i$, and $g_1$ and $g_2$ correspond to the real and imaginary perturbation components, respectively. 
Linearization of Eq.~(\ref{eq1}) about this expression generates the following eigenvalue problem:
\begin{eqnarray}
	\begin{bmatrix}  M_{11}&M_{12} \\  M_{21}&M_{22}\end{bmatrix}\begin{bmatrix} g_1\\ g_2\end{bmatrix}=\lambda\begin{bmatrix} g_1\\ g_2\end{bmatrix} .
\end{eqnarray}
In this linearized operator matrix,
$M_{11}=-[2\sigma\phi _r\phi _i+\kappa (3\phi_r^2+\phi_i^2)-V_i]$,
$M_{12}=-\frac{1}{2}(\frac{\partial^2 }{\partial x^2}+\frac{\partial^2 }{\partial y^2})-V_r+\beta -[\sigma (3\phi_i^2+\phi_r^2)+2\kappa \phi _r\phi _i]$,
$M_{21}=+\frac{1}{2}(\frac{\partial^2 }{\partial x^2}+\frac{\partial^2 }{\partial y^2})+V_r-\beta +[\sigma (3\phi_r^2+\phi_i^2)-2\kappa \phi _r\phi _i]$, and
$M_{22}=+[2\sigma\phi _r\phi _i-\kappa (3\phi_i^2+\phi_r^2)+V_i]$.
The eigenvalue spectrum determines the stability of vortex solitons, and $\lambda$ quantifies the perturbation growth rates. Specifically, the solitons are linearly stable if and only if the dominant eigenvalue satisfies $\lambda_{rm}=\max[\lambda_r] \leq 0$, with instability emerging when $\lambda_{rm}>0$.

The stability of dissipative vortex solitons with distinct topological charges in optical waveguide arrays has revealed in Fig.~\ref{fig4}.
We can find: (i) In azimuthally modulated dissipative structures, vortex solitons with higher topological charge $m$ exhibit markedly enhanced stability (high-$m$ vortex solitons possess well-defined stability windows, while their low-$m$ counterparts remain unstable in their entire existence domains) [Figs.~\ref{fig4}(a) and \ref{fig4}(b)]. \textcolor{black}{This stability contrast is also linked to the field distributions of vortex states [Figs.~\ref{fig2}(f-h)]. Low-charge vortices exhibit several nearly in-phase peaks across neighboring sites, which enhances coupling during propagation and facilitates their conversion into vortices of different charges. By contrast, high-charge vortices develop staggered, out-of-phase patterns that suppress such coupling, rendering them more robust under the balance of localized gain and nonlinear loss.}
(ii) At a relatively low gain level (e.g., $p_i=1.5$), vortex solitons with smaller topological charges $m$ display significantly larger instability growth rates (quantified by $\lambda_{rm}$), leading to shorter propagation distances compared with high-$m$ states. In contrast, at higher gain levels ($p_i\in [4,6]$), the dependence of $\lambda_{rm}$ on $m$ becomes non-monotonic: for $n=20$, $\lambda_{rm}$ decreases with increasing $m$ in the range $m\in [1,5]$, but increases again for $m\in [6,9]$ [Figs.~\ref{fig4}(a) and \ref{fig4}(b)].
\textcolor{black}{It should be noted that for vortices with low topological charges ($m \in [1,4]$), the perturbation growth rates initially decrease with increasing $p_{i}$ for $p_{i}>3$, but subsequently rise beyond a certain threshold and never vanish. For clarity, this latter behavior is not displayed in Fig.~\ref{fig4}(a).}
(iii) For stable vortices, the existence domain progressively narrows with increasing $m$, yet maintains a robust stability region (e.g., $m\in[7,9]$).
Figs. \ref{fig4}(c) and \ref{fig4}(d) show the eigenvalue spectra of representative stable and unstable vortex solitons with charge $m=9$, respectively. Here, the stable vortex soliton exhibits a maximum instability growth rate of $\lambda_{rm}=0$, while the unstable counterpart shows $\lambda_{rm}=0.00565$.
The observed spectral asymmetry--a hallmark of dissipative systems--contrasts sharply with the eigenvalue distributions of conservative and $PT$-symmetric systems.
Notably, these results are obtained for a waveguide array with $n=20$, and similar stability characteristics persist for other values of $n$.

To further validate the results of linear stability analysis, we performed propagation simulations for all vortex solitons. The input optical field was configured as: $\Psi(x,y,z=0)=\phi(x,y)$, implemented with absorbing boundary conditions. The propagation results show excellent agreement with the linear stability analysis. Representative propagation dynamics of vortex solitons with different topological charges are presented in Fig.~\ref{fig5}.
Unstable vortex solitons exhibit dynamic amplitude evolution over a short propagation distance before stabilizing to a constant value [Figs. \ref{fig5}(a1) and \ref{fig5}(b1)].
This behavior suggests the transformation of unstable vortex solitons into solitons with different topological charges.
As evident from Figs. \ref{fig5}(a2) and \ref{fig5}(b2), unstable vortex solitons with different topological charges both evolve into lower-power vortex solitons with a topological charge of $m=9$. 
This optical characteristic reveals the general tendency of unstable vortex solitons to relax into lower-power stable configurations.
For a small $p_i$, unstable vortex solitons with higher topological charges demonstrate markedly extended propagation distances prior to amplitude modulation when compared to their lower-charge counterparts [Figs. \ref{fig5}(a1), \ref{fig5}(b1), \ref{fig5}(a2) and \ref{fig5}(b2)]. 
This observation further corroborates the predictions from the unstable eigenvalue analysis presented earlier.
\textcolor{black}{Furthermore, when the localized gain amplitude becomes sufficiently large, symmetric dissipative vortex solitons carrying higher powers are generally unstable and, after long-distance propagation, evolve into asymmetric nonlinear states characterized by unequal intensities across the waveguides on the ring.}
Stable vortex solitons with both higher- and lower-order topological charges can propagate over considerable distances while maintaining constant amplitude and phase profiles [see rows 3 and 4 of Fig.~\ref{fig5}].
It should be noted that the stable vortex solitons with higher-order and lower-order topological charge do not share the same waveguide array. 
Vortex solitons with lower-order topological charges map to their high-order counterparts in waveguide arrays with reduced waveguide numbers.

Finally, we demonstrate nonlinear excitation of dissipative vortex solitons with distinct topological charges in waveguide arrays of varying channel numbers. 
The input optical field is an annular phase-modulated beam, given by \textcolor{black}{$\Psi(x,y,z=0)=A_{in}\exp[-(\sqrt{x^2+y^2}-r_0)^2/w^2+im\varphi]$,} where $A_{in}$ is the amplitude of the input beam, $r_0$ denotes the radius of the ring-shaped waveguide, \textcolor{black}{$w$ is the waveguide width,} and $\varphi$ is its azimuthal phase.
Lower-$m$ dissipative vortex solitons can be excited in arrays with fewer waveguides [Fig.~\ref{fig6}(a)], while their higher-$m$ counterparts require arrays with more waveguides [Figs. \ref{fig6}(c) and \ref{fig6}(d)]. \textcolor{black}{Stable high-charge vortex solitons can also be excited by Gaussian input beams carrying a phase, even when the beam intensity is weakly distributed across the ring-shaped waveguide array [Fig.~\ref{fig6}(e)].}
Successfully excited dissipative vortex solitons typically exhibit stability in both moderate- and low-power regimes, corresponding respectively to moderate and weak gain levels. 
In contrast, unstable configurations prove difficult to excite [Fig.~\ref{fig6}(b)], with the final output being a lower-power vortex soliton with $m=9$.
These results further validate our previous instability analysis and propagation dynamics simulations.

\section{Conclusion}\label{sec4}
In azimuthally modulated waveguide arrays with linear localized gain and nonlinear loss, we have investigated the existence and stability of dissipative vortex solitons with different topological charges. We find that these solitons are uniformly confined within the waveguide array and fall inside the regions of linear localized gain, while remaining distinguishable by their intrinsic topological phase. The maximum supported charge is constrained by the number of waveguides, implying that higher-charge vortices require arrays with larger azimuthal symmetry. For a given array size, the power and propagation constant of vortex solitons increase monotonically with the level of localized gain. At a fixed gain parameter, however, increasing the topological charge leads to higher field amplitudes, reduced power, and progressively narrower radial confinement. These features become more pronounced in arrays with larger numbers of waveguides. Stable dissipative vortex solitons maintain their field moduli and phase profiles over long propagation distances, whereas unstable ones typically evolve into lower-power vortices of higher charge. Notably, higher-charge vortex solitons exhibit enhanced robustness, and in a $C_{20}$-symmetric array we identified stable dissipative vortices with charges up to $m=6 \sim 9$. The excitation dynamics of the input beams are consistent with the propagation dynamics of the vortex solitons. 
These results provide new insights into the formation and stabilization of high-charge dissipative vortex solitons in structured gain-loss landscapes.

\section*{Declaration of Competing Interest}
The authors declare that they have no known competing financial interests or personal relationships that could have appeared to influence the work reported in this paper.

\section*{Acknowledgements}
This work was supported by the Applied Basic Research Program of Shanxi Province (202303021211191), the National Natural Science Foundation of China (No. 12404385) and China Postdoctoral Science Foundation (No. BX20230218, No. 2024M751950).

\appendix

\end{document}